\documentclass[9pt,twocolumn,twoside]{opticajnl}
\journal{opticajournal} 

\setboolean{shortarticle}{true}

\usepackage{upgreek}
\usepackage{lineno}

\title{300 $\boldsymbol{\upmu}$s optical cavity storage time and $\mathbf{10^{-7}}$ active RAM cancellation for $\mathbf{10^{-19}}$ laser frequency stabilisation}

\author[1]{Adam L. Parke}
\author[1,*]{Marco Schioppo}

\affil[1]{National Physical Laboratory (NPL), Teddington, TW11 0LW, UK}

\affil[*]{marco.schioppo@npl.co.uk}

\begin{abstract}
Frequency stabilisation of lasers to optical reference cavities is an established method to achieve state-of-the-art stability. The strengths of this method are the high discriminator coefficient of optical cavities, and the low-noise extraction of the stabilisation signal using modulation techniques. In this Letter we report beyond state-of-the-art performance on both of these fundamentals, unlocking $\mathbf{10^{-19}}$ fractional frequency laser stabilisation. We employ a $\mathbf{68}\,$cm long cavity to realise an optical storage time of 300 $\boldsymbol{\upmu}$s, achieving ultrahigh frequency discrimination. We develop a simple and robust scheme to actively cancel residual amplitude modulation (RAM) at the $\mathbf{10^{-7}}$ level in an annealed-proton-exchanged lithium-niobate waveguide electro-optic-modulator (EOM).
\end{abstract}

\setboolean{displaycopyright}{false} 

\begin{document}
\maketitle
Cavity-stabilised lasers provide state-of-the-art instability, with recent demonstrations now below the $10^{-16}$ fractional frequency level \cite{Hafner:15, Matei:17, Robinson:19, Schioppo:22}, which makes them key tools of modern optical frequency metrology, with applications ranging from optical atomic clocks \cite{Ludlow:15} to high precision spectroscopy \cite{Argence:15} and gravitational wave detection \cite{LIGO-Virgo-KAGRA:20}. Such performance is derived by the length stability of the optical reference cavities and crucially by the fidelity of the stabilisation technique used to transfer fractional cavity-length variation into laser fractional frequency fluctuation. Modulation methods have been developed to achieve the needed high fidelity so that the information for the stabilisation is encoded at a modulation frequency and $1/f$ and intensity noise can be suppressed. Among them, the Pound-Drever-Hall (PDH) technique \cite{Drever:83, Black:01} has been widely used as it relies on phase modulation, which can be realised with low technical noise, enabling stabilisation at the photon-shot-noise limit of the photodetector (PD). With cavity-stabilised lasers now capable of frequency instability below $10^{-16}$, small but finite technical noise introduced by phase modulators needs to be taken into account. In a phase modulator, a ferroelectric crystal (typically lithium niobate) is used to induce a modulation of the refractive index when an electric RF field is applied (Pockels electro-optic effect). In practice a number of effects in the EOM concurrently play into producing an undesired residual amplitude modulation (RAM) \cite{Wong:85, Li:12}, which depends on the polarisation of the light, degrading the overall fidelity of the stabilisation and therefore of the laser stability. The susceptibility to RAM also depends on the frequency discriminator coefficient of the optical cavity, as summarised by 
\begin{equation}
\frac{\Delta f_{\text{RAM}}}{f_{\text{L}}} = \frac{\Delta V_{\text{PDH RAM}}}{V_{\text{PDH PP}}}\times\frac{1}{\mathcal{D}}\:\:\:\text{with}\:\:\:\mathcal{D}\propto\tau_{\text{cav}}\,, 
\label{eq:1}
\end{equation}
where $\Delta f_{\text{RAM}}$ is the RAM-induced frequency deviation imposed on the laser (with frequency $f_{\text{L}}$), $\Delta V_{\text{PDH RAM}}$ the RAM-induced voltage variation of the PDH error signal, $V_{\text{PDH PP}}$ the voltage amplitude of the PDH error signal peak-to-peak, and $\mathcal{D}$ the optical cavity normalised discriminator coefficient. Hence, long optical storage time reduces the impact of RAM-induced instability of the PDH error signal on the laser frequency. Previous work focused on reducing the RAM of the EOM, with passive \cite{Li:16, Shi:18, Jin:21, Bi:19} or active \cite{Wong:85, Li:12, Gillot:22, Li:14, Zhang:14} methods and with marginal consideration for maximisation of the cavity optical storage time. The lowest levels of RAM were achieved with active RAM compensation implemented on free-space lithium-niobate based modulators \cite{Gillot:22} and in fibre-coupled Ti-indiffused waveguides \cite{Zhang:14}. EOM-free \cite{Kedar:24, Diorico:24} and fibered \cite{Descampeaux:21} implementations were also proposed. 

\begin{figure*}[ht!]
\centering
\includegraphics[width=1\linewidth]{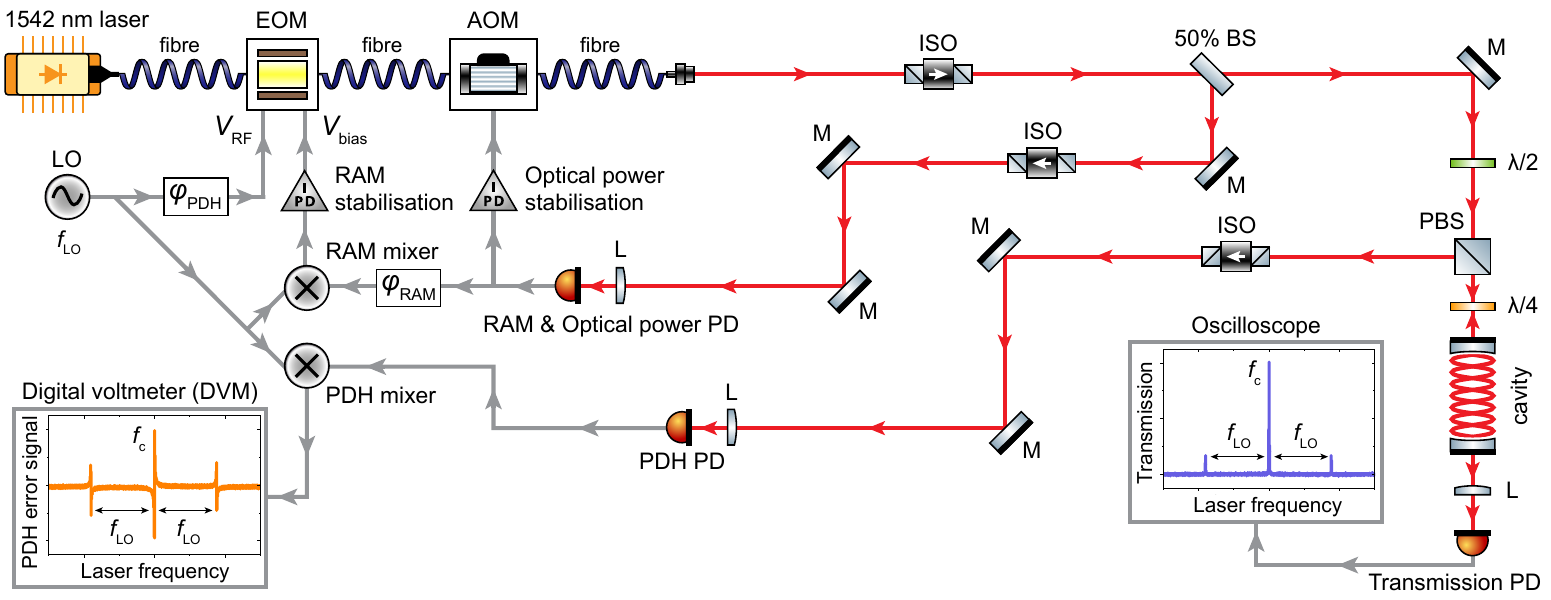}
\caption{Experimental scheme for active RAM cancellation and characterisation. The red (grey) lines represent the free-space optical path (electronic connections) and fibre connections are labeled "fibre". $V_{\text{RF}}$ ($V_{\text{bias}}$) is the RF modulation signal (bias voltage for RAM stabilisation) applied to the EOM. M are mirrors, ISO are free-space optical isolators, L are lenses, PBS is a polarising beam splitter, BS is a beam splitter, PD are photodetectors, $\lambda/4$ is a quarter-wave plate and $\lambda/2$ is a half-wave plate. ${\varphi}_{\text{PDH}}$ and ${\varphi}_{\text{RAM}}$ are passive RF delay lines to fine tune the phase. The plot in orange shows the extracted PDH error signal and the one in blue shows the signal from the transmission PD, both as a function of the laser frequency as it sweeps across the cavity resonance.}
\label{fig:1}
\end{figure*}

In this Letter we focus on reducing the RAM-induced fractional frequency deviation $\Delta f_{\text{RAM}}/f_{\text{L}}$ by actively canceling the RAM at its source (the EOM) and specialising the optical cavity design to enhance the optical storage time. Our approach to mitigating the RAM effect at its source can be summarised as follows: (i) reduce the RAM spatial inhomogeneity, (ii) limit etalon effects, (iii) compensate for the birefringence of the EOM crystal. We address these points by employing a fibre coupled, angled-cut, annealed-proton-exchanged (APE) and x-cut lithium-niobate waveguide EOM (Jenoptik PM1550b100 \cite{Endorsement}). Fibre coupling provides spatial homogenisation of the RAM through propagation inside the fibre. This makes waveguide EOMs more suitable for large beam size cavities and the alignment on PDH and RAM PDs less critical. The angled-cut waveguide provides attenuation of etalon effects \cite{Wooten:00}. Contrary to the Ti-indiffused type, the APE waveguide only lets through one polarization \cite{Wooten:00}, suppressing the co-existence in the waveguide of two polarisations that, experiencing different phase-shifts, can generate RAM downstream. The x-cut crystal ensures a uniform RF field in the waveguide \cite{Wooten:00}. The use of a waveguide EOM enables us to compensate for the birefringence of the EOM crystal with a phase shift obtained by applying an external low voltage bias field ($V_{\text{bias}}$ $\simeq 1\,$V). The experimental setup for this active RAM cancellation scheme is shown in Fig. \ref{fig:1}. Light from a $1542\,$nm laser  is fibre coupled into the waveguide EOM, which is followed by a fibre coupled acousto-optic-modulator (AOM) that is employed for optical power stabilisation and optical isolation. From the AOM, light is fibre out-coupled into free space, encountering an additional stage of optical isolation and a $50\,\%$ power beam splitter, so that two optical branches are produced, one for stabilisation of optical power and RAM, and the other for PDH locking and out-of-loop RAM measurement. Etalon effects arising from spurious reflections from the PDs are suppressed by optical isolators placed on each branch. An RF local oscillator (LO) synthesiser with $f_{\text{LO}}=20\,$MHz drives the EOM. RF branches at $f_{\text{LO}}$ are sent to two frequency mixers, to extract the PDH and RAM stabilisation error signals, respectively. A third PD is placed at the transmission of the optical cavity to measure the intracavity storage time and modulation depth. The latter is estimated from the amplitude ratio between the modulation sidebands and carrier frequency $f_{\text{c}}$ and is set at $15\,\%$, as a trade-off value to generate a sufficiently large PDH error signal and to limit the magnitude of sidebands at higher $f_{\text{LO}}$ harmonics. Passive RF delay lines are used to tune the phases $\varphi_{\text{PDH}}$ and $\varphi_{\text{RAM}}$ to maximise the PDH error signal peak-to-peak and the effectiveness of the RAM cancellation, respectively. The error signal generated by the RAM mixer is processed by a loop filter that closes the feedback on the bias-field port of the EOM, with a bandwidth of $\sim0.5\,$MHz. The out-of-loop RAM $\Delta V_{\text{PDH RAM}}$ and the PDH error signal peak-to-peak $V_{\text{PDH PP}}$ are measured with a digital voltmeter (DVM, Keithley DMM6500 \cite{Endorsement}). 

Equation (\ref{eq:1}) informs us that to convert $\Delta V_{\text{PDH RAM}}/V_{\text{PDH PP}}$ into $\Delta f_{\text{RAM}}/f_{\text{L}}$ we need to know the value of the cavity's normalised discriminator coefficient $\mathcal{D}$ and that the latter is proportional to the cavity optical storage time $\tau_{\text{cav}}$. For a symmetric optical cavity we have $\tau_{\text{cav}}=L_{\text{cav}}/(c\,A)$, where $c$ is the speed of light in vacuum and $A$ is the single mirror total optical power loss.
\begin{figure*}[htb]
\centering
\includegraphics[width=1\textwidth]{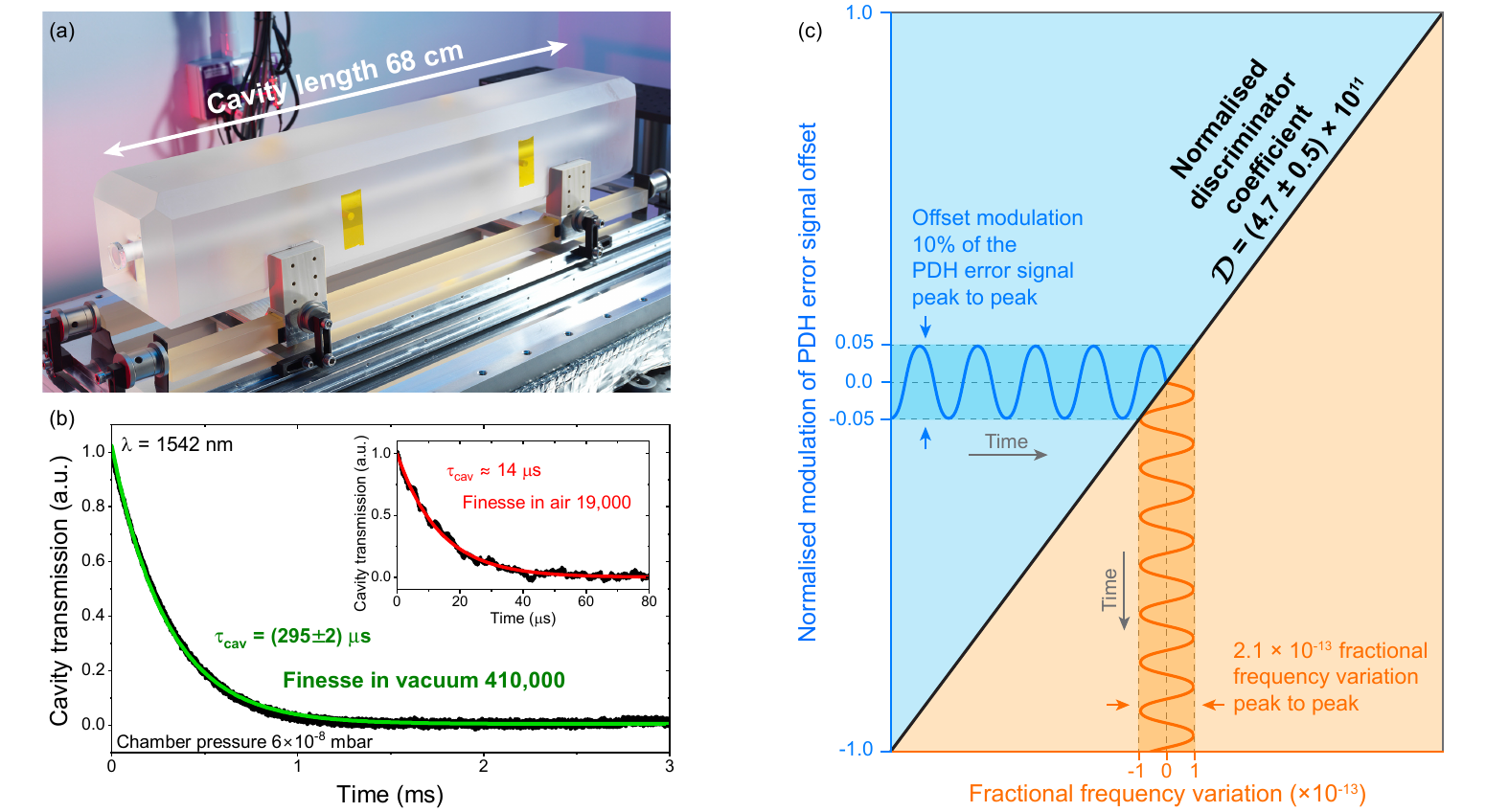}
\caption{(a) Photograph of the 68 cm long cavity used in this Letter. (b) Black points are ringdown measurements in vacuum and in air (in inset plot). The green (red) line is an exponential fit to the ringdown measurement in vacuum (air). (c) Diagram representing the measurement method of the discriminator coefficient $\mathcal{D}$. The blue part of the diagram shows a modulation of the PDH error signal offset at 10 percent of the PDH error signal amplitude peak to peak. The orange part of the diagram shows the resultant fractional frequency variation. The black diagonal line separating the blue and orange parts of the diagram represents the normalised discriminator coefficient, given by the ratio between the PDH offset modulation and the resulting fractional frequency variation.}
\label{fig:2}
\end{figure*}
In this Letter we maximise $\tau_{\text{cav}}$ by developing an optical reference cavity with a length of $68\,$cm, as shown in Fig. \ref{fig:2}(a). A typical ringdown measurement taken in vacuum, as displayed in Fig. \ref{fig:2}(b), gives $\tau_{\text{cav}}=(295\pm2)\,\upmu$s, where the value and the uncertainty are respectively the mean and standard deviation of ten measurements. This corresponds to a cavity finesse $F=\pi\,\tau_{\text{cav}}\,c\,/L_{\text{cav}}\simeq410,000$ and a cavity linewidth of $\Gamma=1/(2\pi\tau_{\text{cav}})\simeq540\,$Hz. This is achieved using dielectric high-reflectivity coatings with $A=7.7\,$ppm over a $1/e^2$ optical beam diameter on the mirror of $1.9\,$mm. This cavity has an estimated thermal noise limit of $2\times10^{-17}$ fractional frequency.
To the best of our knowledge this is the longest optical storage time ever reported for an optical reference cavity with state-of-art thermal noise limit \cite{Zhang:14, Hafner:15, Matei:17, Schioppo:17, Robinson:19, Schioppo:22}. We observed a significant reduction of $\tau_{\text{cav}}$ when measured in air, likely due to water absorption. 

We found some inconsistency in the literature \cite{Gillot:22,Zhang:14,Shi:18} on how to derive $\mathcal{D}$ from a known $\tau_{\text{cav}}$, therefore we opted for a direct measurement. As schematically described in Fig. \ref{fig:2}(c) we add a known voltage modulation to the offset of the PDH error signal ($10\,\%$ of $V_{\text{PDH PP}}$ so that we stay in the linear regime) and we measure the corresponding frequency deviation of the stabilised laser with respect to another reference laser. With this direct method we measure $\mathcal{D}=(4.7\pm0.5)\times 10^{11}$. This value is in good agreement with the expression $\mathcal{D}=f_{\text{L}}/\Gamma$ in \cite{Zhang:14}. 

\begin{figure*}[htb]
\centering
\includegraphics[width=1\textwidth]{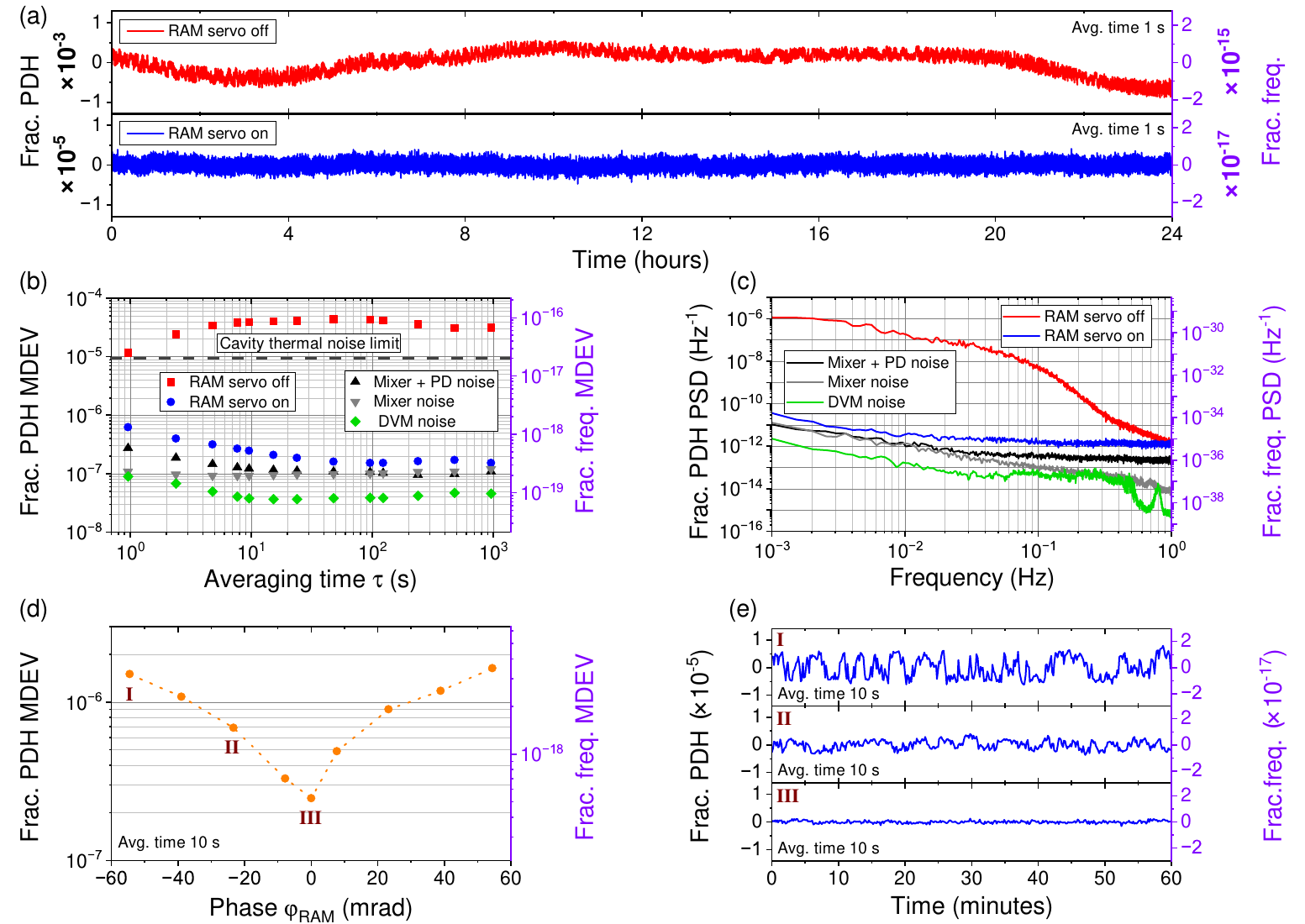}
\caption{(a) Time traces of the PDH error signal, with RAM cancellation off (red line) and on (blue line). (b) and (c) are respectively, the corresponding modified Allan deviations (MDEV) and power spectral densities of the PDH error signal measured in different conditions: with RAM cancellation off; with RAM cancellation on; with no light on the PDH PD (Mixer + PD noise); with the PDH mixer terminated at $50\,\Omega$ (Mixer noise) and with the DVM terminated at $50\,\Omega$ (DVM noise). (d) MDEV of the PDH error signal with RAM stabilisation engaged, as a function of the phase offset between the error signals from the PDH and RAM PDs, represented in \text{Fig. \ref{fig:1}} as ${\varphi}_{RAM}$. (e) Time traces of the PDH error signal corresponding to the points labeled I, II and III on (d). The error bars for the MDEV plots are within the size of the data markers. Linear drift is removed in all plots.}
\label{fig:3}
\end{figure*}

Traditionally, the effects of RAM have been assessed by monitoring the PDH error signal with the laser tuned far from cavity resonances \cite{Li:16, Shi:18, Jin:21, Li:14, Zhang:14, Kedar:24}. As cavity lengths are extended (to reduce the thermal noise limit), the cavity resonances get closer in frequency, increasing the risk of contaminating the RAM measurements with spurious variations of the error signal due to excitation of higher-order modes. To avoid this, we use an open test cavity and place a beam stopper between the cavity mirrors, so that we can eliminate cavity resonances. This enables us to extract a pure RAM effect, which is an upper limit with respect to an unblocked cavity (the modulus of the reflection coefficient of a cavity near resonance is smaller than that of a mirror \cite{Black:01}). In Fig. \ref{fig:3}(a) we show a time trace of $\Delta V_{\text{PDH RAM}}/V_{\text{PDH PP}}$ and the corresponding $\Delta f_{\text{RAM}}/f_{\text{L}}$ with and without active RAM cancellation, over one day of continuous measurement. We demonstrate that active cancellation can reduce the RAM-induced deviations in the error signal voltage and laser frequency by more than two orders of magnitude. We found that the most effective RAM cancellation was achieved without EOM temperature stabilisation or RAM-servo-derived feedback on temperature \cite{Zhang:14}. As shown in Fig. \ref{fig:1} and Fig. \ref{fig:3}(a), with a simpler in-phase stabilisation loop acting only on the bias field of the modulator, it is possible to effectively stabilise the RAM and maintain this stabilisation indefinitely (with a laboratory temperature stability of $\sim1\,^{\circ}$C). Figure \ref{fig:3}(b) shows that the RAM-induced instability without active cancellation would be above the thermal noise limit of our cavity. With active cancellation the RAM-induced fractional voltage instability in the PDH error signal ($\Delta V_{\text{PDH RAM}}/V_{\text{PDH PP}}$) is $6\times10^{-7}$, $2.5\times10^{-7}$ and $1.5\times10^{-7}$, at $1\,$s, $10\,$s and $10-100\,$s integration times, respectively. This corresponds to a RAM-induced fractional instability of the laser frequency ($\Delta f_{\text{RAM}}/f_{\text{L}}$) of $1.3\times10^{-18}$, $5\times10^{-19}$ and $3\times10^{-19}$, at $1\,$s, $10\,$s and $10-100\,$s integration times, respectively, almost two orders of magnitude below our cavity thermal noise limit. To the best of our knowledge, these values are the lowest ever reported \cite{Wong:85, Li:12, Li:14, Zhang:14, Li:16, Bi:19, Jin:21, Gillot:22, Kedar:24, Diorico:24, Shi:18}. Figure \ref{fig:3}(c) displays the noise information in terms of fractional frequency power spectral density, showing that active cancellation reduces noise from $\sim10^{-30}\,{\text{Hz}}^{-1}$ to $\sim10^{-35}\,{\text{Hz}}^{-1}$ at $10\,$s integration time. In Fig. \ref{fig:3}(b) and (c) we include the measurements of the main contributors of technical noise, namely the noise of the DVM, and RF pick-up and ground noise coming from the combination of the PDH mixer and PDH PD, or exclusively from the PDH mixer. For the PDH and RAM error signal extraction we use the same model of mixer (Minicircuits ZRPD-1+ \cite{Endorsement}) and PD (New Focus 1811-FS-AC \cite{Endorsement}). The technical noise measurements suggest that a lower noise PD could lead to improvements from $1\,$s to $10\,$s integration time. Above $10\,$s RF pick-up and ground noise could be tackled by fully digitising the error signal extraction process soon after the photodetection. We found that the effectiveness of the RAM cancellation crucially depends on the phase $\varphi_{\text{RAM}}$, see Fig. \ref{fig:1}, Fig. \ref{fig:3}(d) and (e). The condition for effective RAM cancellation is tight. A phase deviation of $10\,$mrad at $20\,$MHz  (corresponding to a time or delay line length variation of $\sim60\,$ps, or $\sim1\,$cm, respectively) is sufficient to degrade the cancellation, in $V_{\text{PDH RAM}}/V_{\text{PDH PP}}$, from  $2.5\times10^{-7}$ to $5\times10^{-7}$, at $10\,$s integration time. 

In conclusion, we showed that by optimising the cavity design for low RAM susceptibility and actively canceling the RAM, we reduced the RAM-induced frequency deviations to the $10^{-19}$ fractional level. We implemented a $68\,$cm long cavity, achieving an optical storage time of $(295\pm2)\,\upmu$s. We measured the cavity discriminator coefficient. We performed active RAM cancellation with an in-phase feedback loop on an APE waveguide EOM at the $10^{-7}$ level. The demonstrated simplicity, robustness and effectiveness of this technique has the potential for a wide application. We highlighted the importance of phase control to enhance RAM cancellation and we discussed residual sources of technical noise and possible strategies to overcome them.      
\begin{backmatter}
\bmsection{Funding} UK Government Department for Science, Innovation and Technology through UK National Quantum Technologies Programme.
\bmsection{Disclosures} The authors declare no conflicts of interest.
\bmsection{Data availability} Data underlying the results presented in this paper may be obtained from the authors upon reasonable request.
\end{backmatter}
\bibliography{sample}

\begin{thebibliography}{10}
\bibitem{Endorsement}
Commercial products are identified to specify the experimental procedure. This does not imply recommendation or endorsement by NPL.
\end{thebibliography}
\bibliographyfullrefs{sample}
\end{document}